\documentclass[runningheads]{llncs}
\usepackage{graphicx}
\usepackage{float}
\usepackage{color}
\usepackage{geometry}
\usepackage{multirow} 
\usepackage{multicol} 
\usepackage{arydshln}
\usepackage{natbib}
\usepackage{caption}
\usepackage{graphicx, subfig}
\usepackage{booktabs}
\usepackage[font=small,labelfont=bf,labelsep=none]{caption}
\begin{document}

\title{Prediction Model Based on Integrated Political Economy System: \\ The Case of US Presidential Election}
%
%\titlerunning{Abbreviated paper title}
% If the paper title is too long for the running head, you can set
% an abbreviated paper title here
%
%\author{}
\author{Lingbo Li \and
	Ying Fan\thanks{Corresponding author at: No.19 Xinjiekouwai Street, 100875 Beijing, China. Tel: +86 01058807066. \textit{Email addresses: yfan@bnu.edu.cn(Y. Fan), lbLi@mail.bnu.edu.cn (L. Li).} } \and
	An Zeng \and
	Zengru Di}
%
%\authorrunning{ }
%\institute{}
\authorrunning{ }
% First names are abbreviated in the running head.
% If there are more than two authors, 'et al.' is used.
%
%\institute{}
\institute{\textit{School of Systems Science, Beijing Normal University, 100875, Beijing, China}}
\maketitle              % typeset the header of the contribution
\begin{abstract}
This paper studies an integrated system of political and economic systems from a systematic perspective to explore the complex interaction between them, and specially analyzes the case of the US presidential election forecasting. Based on the signed association networks of industrial structure constructed by economic data, our framework simulates the diffusion and evolution of opinions during the election through a kinetic model called the Potts Model. Remarkably, we propose a simple and efficient prediction model for the US presidential election, and meanwhile inspire a new way to model the economic structure. Findings also highlight the close relationship between economic structure and political attitude. Furthermore, the case analysis in terms of network and economy demonstrates the specific features and the interaction between political tendency and industrial structure in a particular period, which is consistent with theories in politics and economics.

\end{abstract}
\section{Introduction}

\quad\  In view of the high complexity of real-world systems, researchers should conduct research from a systematic perspective to better understand the mechanism of systems and control them efficiently, which requires us not only to take into account the multilevel structure within a single system, but also to explore the coupling interaction between systems \citep{england2017households}. For example, if we only focus on the own factors of political system (e.g. campaign slogans and canvassing activities) and ignore the complex interaction existing between political system and other real-world systems (e.g. economics and transportation), the conclusion may fall into an unrealistic misunderstanding. Therefore, researchers attempt to treat the political system and its coupling systems as a more complex whole \citep{levin2016great,rhue2014digital}, which leads to new phenomena that have never been captured in a single network.

Especially in economic science, there is a large and growing literature on the influence of politics on economic outcomes. Government adjusts the industrial structure and keypoint of economic development through administrative means, while the economy, as the combined effect of government regulation and free markets, can influence policy formulation through feedback. \citet{acemoglu2013economics} inform us that the fundamental approach to policy prescription in economics derives from the recognition that market failures creates room for well-designed public interventions. Moreover, governments in the relatively advanced economies are able to raise higher tax revenues and play a more important role in the economy \citep{acemoglu2005politics}. Given the close relationship between politics and economy, there is a need for systematic approaches to specifically analyze the interaction, so we utilize network science to integrate politics and economy into a coupling system to study.

We take the presidential election as an example of typical political events. In the scientific research of election forecasting, there are four main schools: structuralists \citep{lewis2014us}, aggregators \citep{berg2008results}, synthesizers \citep{rothschild2015combining} and judges \citep{cook2014recalibrating}, according to the article published in the journal Science by \citet{kennedy2017improving}. Recent studies of election prediction are largely based on polls \citep{spenkuch2018political,murr2016wisdom} or data from online social platforms, such as twitter \citep{cameron2016can,kagan2015using}. Many cases improved regression models to make better predictions \citep{xie2018big,parackal2018value}, but apart from traditional methods, machine learning has also received much attention due to its capability of processing big data \citep{karami2018mining,sharma2016prediction}. Therefore, more and more researchers utilize machine learning and other big data processing technology to predict elections, e.g. sentiment analysis \citep{jose2016prediction,ramteke2016election} and signal processing \citep{xie2016wisdom}. However, there are some problems to be solved in previous works. For instance, some works are limited to the political field, lack of divergent thinking that society is the result of coordinated operation; some require tedious data collection and complex processing, while the polls and social media data also have bias \citep{shirani2018disentangling,huberty2015can}.

To break through the obstacles, our work aims at propose a framework, which operates efficiently, requires easily accessible data and enables a comprehensive analysis of political and economic systems. Specifically, we exploit basic economic data to establish signed networks, and study the diffusion and evolution of the US states’ attitudes during the election through the dynamics on signed networks, which relies on a kinetic model called the Potts Model, differing from the previous works on opinion diffusion on political networks \citep{gonzalez2014assessing,huckfeldt2014noise}. In this way, we provide a simple and efficient prediction model for the US presidential election, and identify the close relationship between economic structure and political attitudes.          

Signed networks, which can represent both positive and negative relationships by edges with different signs, have been used to describe the clear confrontation and alliance in political system \citep{doreian2013partitioning,smith2014power}, but in this study, they efficiently characterize the underlying economic relationships between political individuals. We use basic economic indicators to calculate the similarity between industrial structures of regions, and construct association networks to reveal the implicit economic relationship between individuals, where the signed network helps to strengthen the comparison of economic structure relations and facilitate the subsequent analysis. In addition, as reported by \citet{schweitzer2009economic} that economic networks are facing new challenges, previous studies on economic networks largely focus on the economic benefits of social networks \citep{fracassi2017corporate} or networking intuitively from economic activity data (e.g. commerce and investment) \citep{aldasoro2017bank,werth2013co}, therefore our framework is a new way to build economic networks.

The rest of this paper is organized as follows. In Section 2, we interpret the feasibility of our framework and how to construct the signed networks. Section 3 and 4 introduce our prediction model and its simulation process, and present the experimental results. Section 5 complements the analysis of network structure and economic characteristics. Section 6 concludes.

\section{Data and Networks}

\subsection{Data analysis}
\quad\ We use basic to analyze the differences of industrial structure between different regions and establish networks, and utilize the preliminary polling data to support the prediction model. In the case of predicting the US presidential election, our economic data only includes the annual Gross Domestic Product (GDP) information of 50 states and Washington DC during twelve years from 2005 to 2016, which is easily available on official US commerce website \textit{(www.bea.gov)}. The GDP data of each state is subdivided into 17 categories according to the economic use, such as construction and retail trade, which can be grouped into three major industries and government revenue.  

\begin{table}[h]  
	\centering
	\caption{The Friedman test on the GDP data of each industry.}  
	\label{tab1}
	\begin{tabular}{p{3.5cm}p{8.5cm}p{1.7cm}p{0.7cm}p{1.7cm}}
		\hline
		\hline
		\multicolumn{5}{c}{\bf\centering Friedman Test} \\
		\hline
		
		\scriptsize  \textbf{Data (2004-2016)} & \scriptsize \textbf{Data details} & \scriptsize  \textbf{Chi-squared} & \scriptsize  \textbf{Df} & \scriptsize  \textbf{P-value} \\
		\hline 
		
		\scriptsize Agriculture Industry & \scriptsize Agriculture, forestry, fishing, and hunting & \scriptsize 44.91 & \scriptsize 4 & \scriptsize 4.151E-09 ***\\
		\cdashline{1-5}[0.8pt/2pt]
		
		\multirow{3}*{\scriptsize Manufacturing Industry} & \scriptsize Mining, quarrying, and oil and gas extraction & \multirow{3}*{\scriptsize103.00} & \multirow{3}*{\scriptsize4} & \multirow{3}*{\scriptsize \textless 2.2E-16 ***} \\
		\cdashline{2-2}[0.8pt/2pt]	
		&\scriptsize Construction & & & \\
		\cdashline{2-2}[0.8pt/2pt]		
		&\scriptsize Manufacturing & & & \\
		\cdashline{1-5}[0.8pt/2pt]		
		
		\multirow{10}*{\scriptsize Service Industry} & \scriptsize Utilities & \multirow{10}*{\scriptsize118.32} & \multirow{10}*{\scriptsize4} & \multirow{10}*{\scriptsize \textless 2.2E-16 ***} \\
		\cdashline{2-2}[0.8pt/2pt]		
		&\scriptsize Wholesale trade & & & \\
		\cdashline{2-2}[0.8pt/2pt]		
		&\scriptsize Retail trade & & & \\
		\cdashline{2-2}[0.8pt/2pt]
		&\scriptsize Transportation and warehousing & & & \\
		\cdashline{2-2}[0.8pt/2pt]
		&\scriptsize Information & & & \\
		\cdashline{2-2}[0.8pt/2pt]
		&\scriptsize Finance, insurance, real estate, rental, and leasing & & & \\
		\cdashline{2-2}[0.8pt/2pt]
		&\scriptsize Professional and business services & & & \\
		\cdashline{2-2}[0.8pt/2pt]
		&\scriptsize Educational services, health care, and social assistance & & & \\
		\cdashline{2-2}[0.8pt/2pt]
		&\scriptsize Arts, entertainment, recreation, accommodation, and food services & & & \\
		\cdashline{2-2}[0.8pt/2pt]
		&\scriptsize Other services (except government and government enterprises) & & & \\
		
		\cdashline{1-5}[0.8pt/2pt]	
		\multirow{3}*{\scriptsize Government Revenue} & \scriptsize Federal civilian & \multirow{3}*{\scriptsize33.91} & \multirow{3}*{\scriptsize4} & \multirow{3}*{\scriptsize 7.762E-07 *** } \\
		\cdashline{2-2}[0.8pt/2pt]		
		&\scriptsize Military & & & \\
		\cdashline{2-2}[0.8pt/2pt]		
		&\scriptsize State and local & & & \\
		
		\hline
		\hline 
	\end{tabular}
\end{table}

It is preliminarily verified by data analysis that using specific GDP data to construct association networks of industrial structure is feasible. In order to make a longitudinal analysis, we first conduct non-parametric test on the differences of the overall American industrial structure between different years. Specifically, these annual GDP data with 17 categories are grouped into four industries, including agriculture, manufacturing, service and government revenue. For each industry, we integrate its percentage in 51 American states, so we can get four sets of data, each containing 51 numbers, in every election year between 2004 and 2016. Therefore, five sets of data for each industry from five different election years are used as paired samples for the Friedman test. According to the result shown in Table 1, there are significant differences in the data of each industry between years, and the results passed the significance test. This reveals that the industrial structure of the American states did change during the period.

\begin{figure}
	\centerline{\includegraphics[scale=0.2]{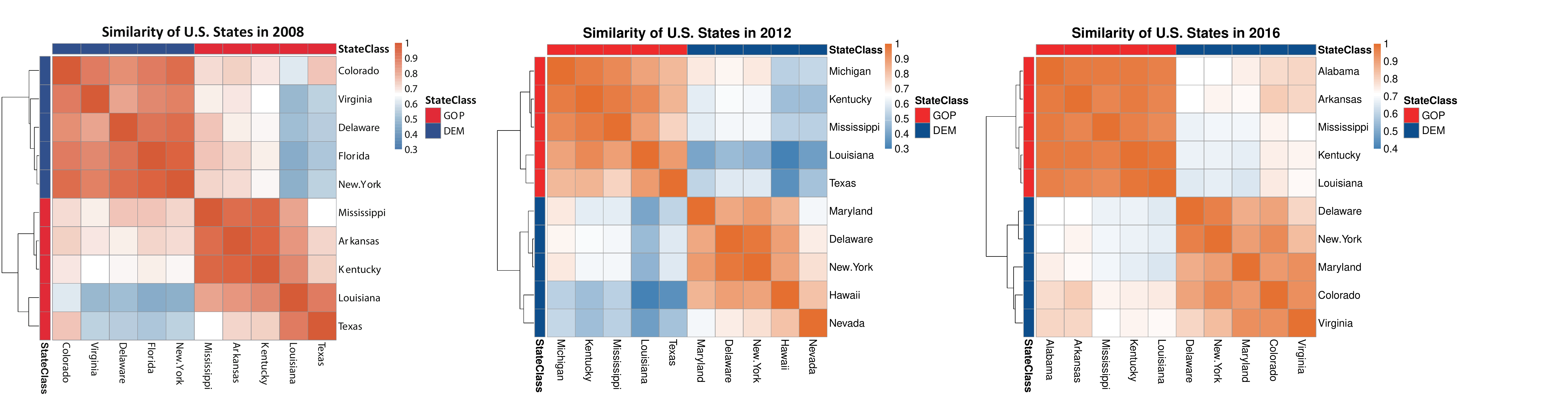}}
	\caption{ Comparison of the modified cosine similarity of industrial structure among five red states and five blue states in each election year from 2008 to 2016.} 
	\label{fig1}
\end{figure}

On the other hand, we analyze the differences of industrial structure between different American states to make a horizontal analysis. It is generally known that in the US presidential election, American states can be roughly grouped into three categories: the red states, the blue states and the swing states. Historically, people in the red states will predominantly vote for the Republican Party and in the blue states for the Democratic Party \citep{you2015multifaceted}. We consider the GDP data of each state as a 17-dimensional vector, and calculate the modified cosine similarity (more details given in chapter 2.2) between these vectors. Especially, we compare the results of several representative red and blue states. Shown in Fig. \ref {fig1}, the similarity between states, calculated with data from every election year between 2008 and 2016, is displayed by heat maps, indicating that the similarity within red states or blue states looks much higher than that between red state and blue state. 

Therefore, based on both the longitudinal and horizontal analysis, we believe that it make sense to show the differences of industrial structure between states and changes of that between years by using GDP data to construct association networks.

\subsection{Network Construction}

\quad\ Given that signed networks, representing both positive and negative relationships simultaneously, can efficiently characterize many real-world systems, ranging from social networks to biological networks, we utilize signed networks to give a clearer picture of the economic similarity between different regions.

The association network of industrial structure for each year consists of 51 nodes, representing the 50 states and Washington DC in the United States, and several weighted edges, representing the similarity of industrial structure between every two states based on annual GDP data. To build such networks, we firstly assign a 17-dimensional vector to each node, recording the GDP data of 17 subcategories in every state, and compute the modified cosine similarity between each two vectors as follows:

\begin{equation}
w_{A, B}^{0}={sim}(\mathrm{A}, \mathrm{B})=\frac{R_{A} \cdot R_{B}}{\left\|R_{A} |\right\|_{R} R_{B} \|}=\frac{\sum_{i}^{n} R_{A, i}^{*} R_{B, i}}{\sqrt{\sum_{i}^{n} R_{A, i}^{2}} * \sqrt{\sum_{i}^{n} R_{B, i}^{2}}}
\end{equation}

\noindent Here, $ R_A $ is the modified vector of node A, $R_{A}=R_{A}^{0}-\overline{R_{A}}$, where $ R_{A}^{0} $ and $\overline{R_{A}}$ represent the vector of raw data and the vector obtained by averaging in each dimension. It is the same for $ R_B $. Since the level of economic development varies greatly among the states, there is large difference between the absolute values of these original vectors, and the modified vectors aims to partially mitigate the interference caused by such huge difference. $ R_{B, i} $ and $ R_{B, i} $ are the values in the ith dimension of the vector of node A and node B respectively. 

\begin{figure}
	\centerline{\includegraphics[scale=0.4]{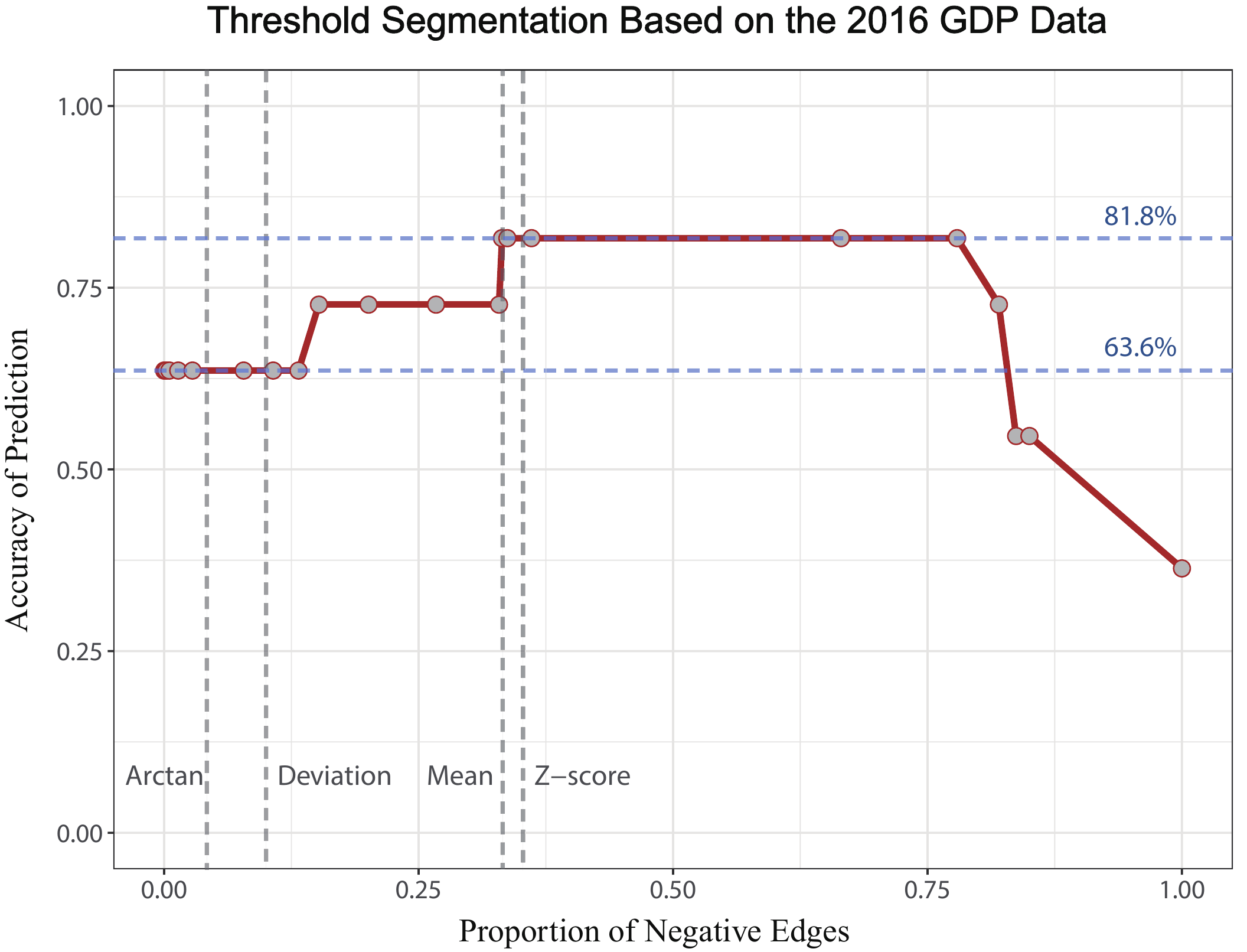}}
	\caption{Relationship between the proportion of negative edges and prediction accuracy. The solid red line shows that the accuracy of prediction changes along with the proportion of negative edges, which is adjusted by threshold segmentation. The grey dotted lines indicate the negative edges ratio of signed networks obtained by four standardization algorithms. These results are based on the 2016 annual US GDP data.} \label{fig2}
\end{figure}

The next step after getting the initial value of similarity $ R_{A,B}^{0} $ is data standardization. As an example, the comparison of four standardization algorithms based on the 2016 annual US GDP data is shown in Fig. \ref{fig2} with grey dotted lines, while the solid red line indicates the accuracy of prediction changes along with the proportion of negative edges in the network, where the negative edges ratio was adjusted by threshold segmentation. As we can see from the graph, with the increase of negative edge ratio in the network, the prediction accuracy increases gradually and then remains stable, while when the proportion of negative edges is too high, the accuracy decreases sharply. In other words, adding negative edges of appropriate proportion to the network can effectively improve the prediction accuracy, and therefore signed networks can play an important role in our work.

We test these four standardization algorithms to process the similarity $w_{A, B}^{0}$, including deviation standardization $\left(\mathrm{x}^{\prime}=\left(x-x_{\min }\right) /\left(x_{\max }-x_{\min }\right)\right)$, mean normalization $\left(\mathrm{x}^{\prime}=(x-\mu) /\left(x_{\max }-x_{\min }\right)\right)$, arctan function normalization $(y={atan}(x) * 2 / \pi)$, and z-score standardization, whose calculation formula is shown in Equatio. 2.

\begin{equation}
w_{A, B}=\frac{w_{A, B}^{0}-\bar{w}}{s t d}
\end{equation}

\noindent Here, the $ \bar{w} $ and $std$ represent the mean and standard deviation of all the similarity values respectively, and $ w_{A, B} $ represents the standardized similarity between vectors of node A and node B, which may be greater than or less than zero and act as the weight of corresponding edge in the network.

According to the overall results based on the data of different years, the z-score standardization, which can maintain the original distribution of data, performs best in getting signed networks of appropriate negative edges ratio among all the four algorithms, so we standardized the similarity values by z-score. As a result, we construct one signed network for every year, which consists of 51 nodes and both positive and negative weighted edges. In these networks, the higher the similarity between the pair of nodes is, the higher the edge weight is.

\section{Our Prediction Model}
\quad\ In previous works, many researchers refined complex real-world systems into network models and conducted research on network models to explore the properties of real-world systems. In this paper, we apply a dynamic model, the Potts Model, to the network to simulate the evolving process of public opinion \citep{li2019binary,zhou2018random}.

\subsection{Kinetic Model: the Potts Model}

\quad\ The Potts Model, as a classical model in statistical physics, is a generalization of the Ising Model to more-than-two components \citep{wu1982potts}. This model is usually placed on two-dimensional rectangular Euclidean lattices, which may be generalized to other dimensions or other lattices  \citep{castaneda2008parallel}. On each site of the lattice, there is a spin that can take q different values ($\sigma_{i}=1,\ldots,q$) (especially $q = 2$ for Ising model) and interaction between spins determine their spin directions according to the Hamiltonian of the system \citep{essam1979potts}. The Hamiltonian function is defined as follows:

\begin{equation}
\mathcal{H}_{Z^{2}, b}(\sigma)=-J \sum_{<i, j>} \delta_{\sigma_{i}, \sigma_{j}}-b \sum_{i} \delta_{\sigma_{i}, 1}
\end{equation}

\noindent where $J$  is a positive coupling constant and $b$ represents the strength of the external field. $\sigma$ is the Kronecker delta ($\delta(\mathrm{x}, \mathrm{y})=0 $ if $ \mathrm{x} \neq \mathrm{y}$  and $ 1 $ if $ \mathrm{x}=\mathrm{y})$ and $\left(\sigma_{i}, \sigma_{j}\right)$ denotes the states of neighbor node pair $(i,j)$. The partition function is defined as:

\begin{equation}
Z_{Z^{2}, b}(\sigma)=\sum_{\sigma} \exp \left\{K \sum_{<i, j>} \delta_{\sigma_{i}, \sigma_{j}}+h \sum_{<i, j>} \delta_{\sigma_{i}, 1}\right\}
\end{equation}

\noindent\ Due to $K=\beta J, h=\beta b$ and $\beta=1 /\left(k_{B} * T\right)$, the above formula can be rewritten as:

\begin{equation}
Z=\sum_{\sigma} \exp \left\{\frac{J}{k_{B} T} \sum_{<i, j>} \delta_{\sigma_{i}, \sigma_{j}}+\frac{b}{k_{B} T} \sum_{<i, j>} \delta_{\sigma_{i}, 1}\right\}=\sum_{\sigma} \exp \left\{\frac{-H}{k_{B} T}\right\}
\end{equation}

\noindent where $k_{B}$ is the Boltzman constant and $T$ is a certain temperature value.

According to this model, the system will evolve in the direction of energy reduction, and whether the new state of a spin is accepted depends on the probability function, which may be the Glauber transition function \citep{glauber1963time} or the Metropolis transition function \citep{metropolis1953equation}. In our work, we use the Metropolis Monte-Carlo simulation to proceed the evolving process, which relies on a switching probability $P$ by writing in

\begin{equation}
P=\left\{\begin{array}{ll}
	{\exp \left\{-\frac{\Delta H}{k_{B} T}\right\}} & { { if } \Delta H>0} \\
	{1} & { { if } \Delta H \leq 0}
\end{array}\right.
\end{equation}

\noindent Where $\Delta H=H_{2}-H_{1}$.

\subsection{Monte-Carlo simulation}

\quad\ In the case of predicting the US presidential election, we apply the Potts Model to weighted signed networks and employ a simple Monte-Carlo heat-bath algorithm with simulated annealing \citep{kirkpatrick1983optimization}.

In these signed networks, the node state is expressed by $S_{i}$ and has five possible values ($q=5$), representing five different political attitude: solidly (dark-red state, $S_{i}=1$) or slightly (light-red state, $S_{i}=2$) supporting the Republican Party, neutral attitude (swing state, $S_{i}=3$), slightly (light-blue state, $S_{i}=4$) or solidly (dark-blue state, $S_{i}=5$) supporting the Democratic Party. The state of a node can only be changed to its adjacent states (i.e. if the state of chosen node is $S_{i}=2$, its next possible state can only be $S_{i}=1$ or $3$), which makes our evolution mechanism more realistic.

In addition, a node’s state is affected by its neighbor nodes according to the Hamiltonian, which is calculated as follows:

\begin{equation}
\mathrm{H}=-\mathrm{J} \sum_{<i, j>} \delta_{\sigma_{i}, \sigma_{j}}=-\sum_{<i, j>} a_{i, j} * \delta_{\sigma_{i}, \sigma_{j}}
\end{equation}

\noindent Here, the coupling constant $J$ is replaced by $a_{i, j}$, which is the weight of edge between node i and node j and also is the value in the location $(i,j)$ of the adjacency matrix of the network, and the Kronecker delta guarantees that $\delta\left(\sigma_{i}, \sigma_{j}\right)=0$ if $\sigma_{i} \neq \sigma_{j}$ and 1 if $\sigma_{i}=\sigma_{j}$.Therefore, the energy constraint includes two cases. If a pair of nodes is connected by positive edge, the same state of nodes will lead to a decrease in the Hamiltonian while the different states will lead to an increase; if negative edge exists, it is opposite. 

To perform simulated annealing, starting from an initial temperature, the system is subsequently cooled down, until it reaches a configuration where the variation range of the Hamilton less than a given value during a certain number of sweeps over the network. The simulation proceeds as follows: 

1) The initial configurations of signed networks are determined by the polls released at the beginning of the US presidential election, and we calculate the Hamiltonian of current system $H_{1}$.

2) Select a node A randomly and then choose one of its adjacent states as its new state, $S_{A^{\prime}}$, with a given probability. The state-transition probability of dark-red (dark-blue) nodes is smaller than that of other nodes, indicating the fact that the stalwarts are less likely to change their attitudes according to \citet{bello2014influence}.

3) Calculate the new Hamiltonian $H_{2}$ after changing the state of node A, and the energy difference $\delta \mathrm{H}=H_{2}-H_{1}$. 

4) Judge whether to accept the state transition of node A based on the probability P defined in formula.5: if the Hamiltonian decreases ($\delta \mathrm{H}<0$), the new state will be accepted; if the Hamiltonian increases ($\delta \mathrm{H}>0$), the new state will be accepted with a small probability $e^{\delta \mathrm{H} / k_{B} T}$.

5) Repeat step 2 to 4 for N times so that we completed a Monte Carlo Step evolution, which is defined as $\Delta t_{M C}=n / N$ \citep{raabe2000scaling}.

6) Repeat step 5 up to $m_{1}$ times until the system evolves to a steady state at the current temperature, where the Hamiltonian fluctuation satisfies a small limited interval. 

7) Continuously lower the experimental temperature $m_{2}$ times according to the formula $T_{k+1}=\alpha * T_{k}$ and repeat step 6 to make the system reach steady state at each temperature. 

8) When the temperature drops to a very low level, the system achieves a final configuration where the Hamiltonian tends to remain unchanged.

\begin{figure}
	\centerline{\includegraphics[scale=0.24]{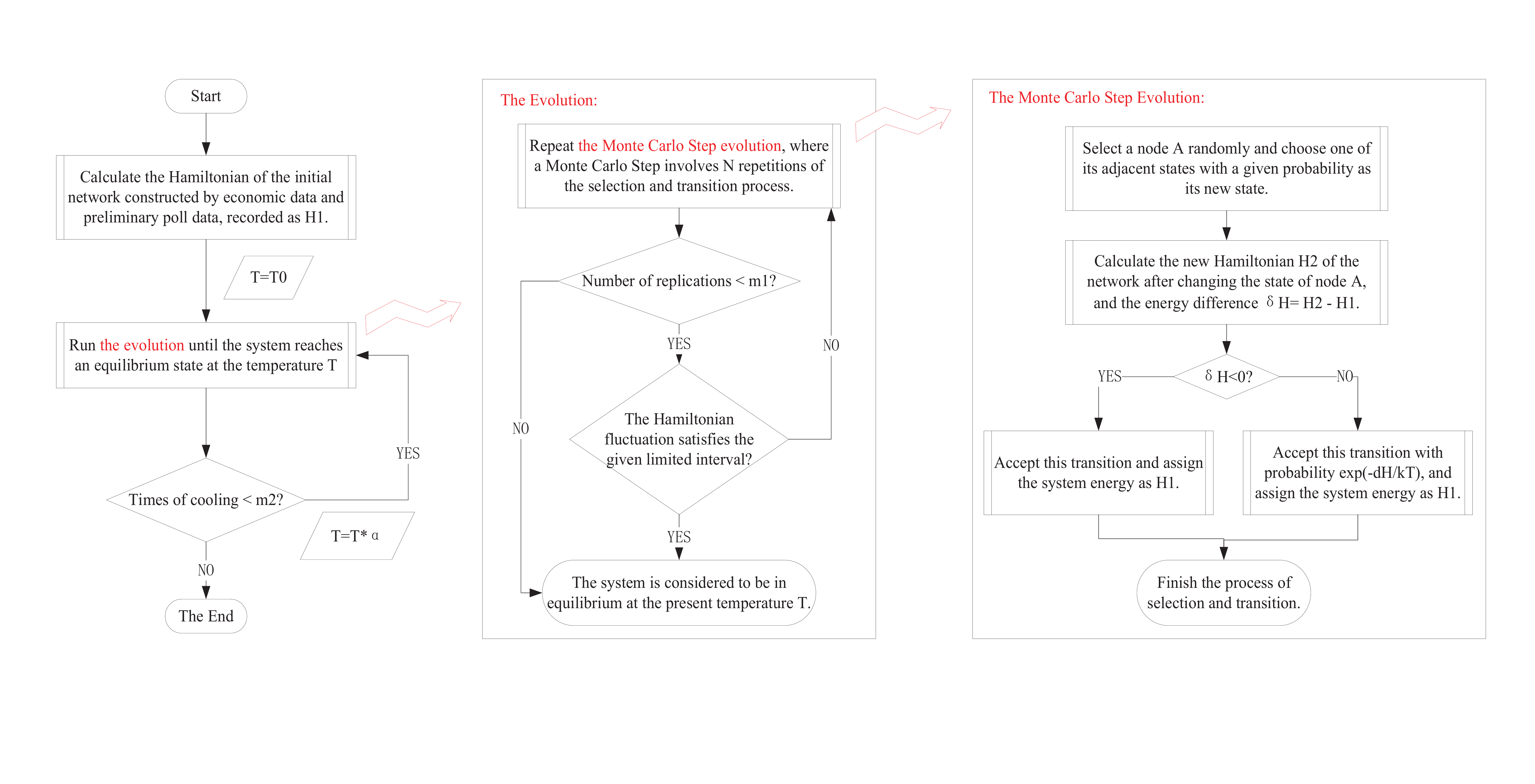}}
	\caption{The Monte Carlo simulation of our dynamic model applied to the US election forecasting based on the Potts Model.} 
	\label{fig3}
\end{figure}

We repeat the procedure described above, which is shown in Fig. \ref{fig3}, at least $m_{3}$ times and take the average as the result to avoid the error caused by outliers. It should be noted that edges of the network do not change during the evolution.

\section{Results}

\quad\ We apply the above experiments to the case of predicting US presidential election, simulating the impact of economic relations among American states on their political attitudes during the presidential election. In our work, the parameter settings are as follows: $\alpha=0.9, \mathrm{N}=51, m_{1}=1000, m_{2}=20,m_{3}=1000$. To select an appropriate $T_{0}$, we assign different values to $k_{B} T$ to observe the forecast results in the US presidential elections between 2008 and 2016, including 0.1, 0.5, 1, 2 and 2.76, whose results are recorded in Table \ref{tab2}. As shown in the table, we record the predicted winning party, the forecast accuracy in swing states and that in other states of the experiments with different parameters, which are calculated from the prediction results of American states. The results indicate that the prediction effect tends to be the best when $k_{B} T=1$, which determines our choice of parameter $T_{0}$.

\begin{table}[h]  
	\caption{The prediction results of experiments with different initial temperature $T_{0}$.}  
	\label{tab2}
	\begin{tabular}{p{4.3cm}p{4cm}p{1.5cm}p{1.5cm}p{1.5cm}p{1.5cm}p{1.5cm}}
		\hline
		\hline
		\scriptsize \centering \textbf{Predicted object} & \multicolumn{6}{c}{\scriptsize \bf\centering Parameter and forecast results}  \\
		\hline
		
		\multirow{3}*{\scriptsize The 2008 US Presidential Election} & \scriptsize $k_{B} T$ & \scriptsize 0.10  & \scriptsize 0.50 &  \textcolor{red}{\scriptsize 1.00} & \scriptsize 2.00 & \scriptsize 2.76 \\
		\cline{2-7}	
		&\scriptsize Predicted winning party & \scriptsize R & \scriptsize D & \textcolor{red}{\scriptsize D} & \scriptsize D & \scriptsize D \\
		\cdashline{2-7}[0.8pt/2pt]
		&\scriptsize Forecast accuracy in swing states & \scriptsize - & \scriptsize 100.0\% & \textcolor{red}{\scriptsize 100.0\%} & \scriptsize 76.9\% & \scriptsize 76.9\% \\
		\cdashline{1-7}[0.8pt/2pt]
		\scriptsize Number of swing states: 13 &\scriptsize Forecast accuracy in other states & \scriptsize - & \scriptsize 97.4\% & \textcolor{red}{\scriptsize 97.4\%} & \scriptsize 76.3\% & \scriptsize 55.3\% \\
		\hline
		
		\multirow{3}*{\scriptsize The 2012 US Presidential Election} & \scriptsize $k_{B} T$ & \scriptsize 0.10  & \scriptsize 0.50 & \textcolor{red}{\scriptsize 1.00} & \scriptsize 2.00 & \scriptsize 2.76 \\
		\cline{2-7}	
		&\scriptsize Predicted winning party & \scriptsize D & \scriptsize D & \textcolor{red}{\scriptsize D} & \scriptsize D & \scriptsize D \\
		\cdashline{2-7}[0.8pt/2pt]
		&\scriptsize Forecast accuracy in swing states &  \scriptsize 55.6\% & \scriptsize 44.4\% & \textcolor{red}{\scriptsize 66.7\%} & \scriptsize 66.7\% & \scriptsize 88.9\% \\
		\cdashline{1-7}[0.8pt/2pt]
		\scriptsize Number of swing states: 9 &\scriptsize Forecast accuracy in other states & \scriptsize 100.0\% & \scriptsize 100.0\% & \textcolor{red}{\scriptsize 100.0\%} & \scriptsize 95.2\% & \scriptsize 59.5\% \\
		\hline
		
		\multirow{3}*{\scriptsize The 2016 US Presidential Election} & \scriptsize $k_{B} T$ & \scriptsize 0.10  & \scriptsize 0.50 & \textcolor{red}{\scriptsize 1.00} & \scriptsize 2.00 & \scriptsize 2.76 \\
		\cline{2-7}	
		&\scriptsize Predicted winning party & \scriptsize R & \scriptsize R & \textcolor{red}{\scriptsize R} & \scriptsize R & \scriptsize R \\
		\cdashline{2-7}[0.8pt/2pt]
		&\scriptsize Forecast accuracy in swing states &  \scriptsize 72.7\% & \scriptsize 72.7\% & \textcolor{red}{\scriptsize 81.8\%} & \scriptsize 81.8\% & \scriptsize 72.7\% \\
		\cdashline{1-7}[0.8pt/2pt]
		\scriptsize Number of swing states: 11 &\scriptsize Forecast accuracy in other states & \scriptsize 100.0\% & \scriptsize 100.0\% & \textcolor{red}{\scriptsize 100.0\%} & \scriptsize 87.5\% & \scriptsize 62.5\% \\
		
		\hline
		\hline 
	\end{tabular}
	
	\textit{Note: in the part of ‘Predicted winning party’, ‘R’ represents the Republican Party and ‘D’ represents the Democratic Party. If the predicted winning party does not match the fact, the forecast accuracy in swing states and that in other states will no longer be calculated. The cases producing the best prediction result are marked in red.}
\end{table}

\begin{figure}
	\centerline{\includegraphics[scale=0.24]{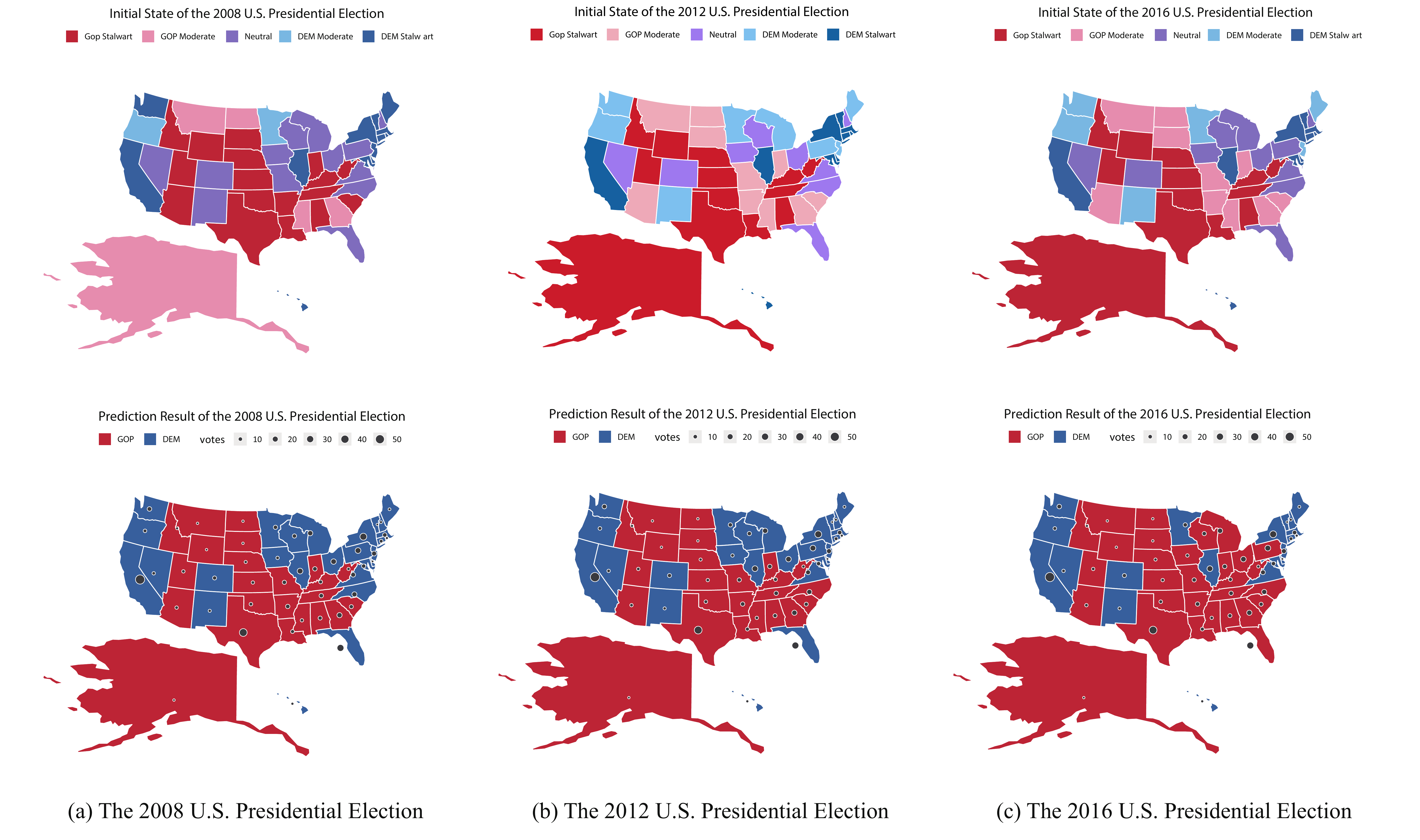}}
	\caption{The comparison of the initial system and the prediction results of the US Presidential Elections from 2008 to 2016, shown in chronological order in (a) \~{} (c). These are the results of using the GDP data of the election year to predict the election result of that year. The gray bubbles in the three figures in the lower part represent the number of votes in each state; the larger the bubble, the more votes there are in that state.} 
	\label{fig4}
\end{figure}

The Fig. \ref{fig4} compares the initial state and the predicted steady state of the US presidential election system, and the Table \ref{tab3} records the detailed prediction results of experiments using networks of different years to predict the different US presidential elections. Based on these results, we have the following findings:
 
1)  We can see from the simulation results shown in Fig. \ref{fig4} that in the last three elections, using the economic data of the election year can accurately simulate the actual election results of that year. This is the evolution results on economic networks; therefore, we believe that this can demonstrate the relationship between economic structure and politics.

2)  According to the detailed results shown in Table \ref{tab3}, it is verified that compared with the economic data of other years, using the network based on data of the election year produce the best prediction result, which is marked in red in Table \ref{tab3}. The best performance of the economic data of the election year confirms the close interaction between the development of industrial structures and political attitudes.

3)  If we use a network based on the economic data of last three years before the election year, the prediction result can succeed in pointing out the winning party. This indicates that our model provides a reliable new idea for the US presidential election.

These findings reveal the relationship between economic development and political tendency. Because of different economic policies and the free market, the industrial structure of each state has been adjusted over time, so that in our networks of different years, there are many changes in the weighted edges between nodes and network structure features, leading to different evolution results.

\begin{table}[h]  
	\caption{The prediction results of experiments using networks of different years to predict the different US presidential elections. }  
	\label{tab3}
	\begin{tabular}{p{4.3cm}p{4cm}p{1.5cm}p{1.5cm}p{1.5cm}p{1.5cm}p{1.5cm}}
		\hline
		\hline
		\scriptsize \centering \textbf{Predicted object} & \multicolumn{6}{c}{\scriptsize \bf\centering Data and forecast results}  \\
		\hline
		
		\multirow{3}*{\scriptsize The 2008 US Presidential Election} & \scriptsize Network & \scriptsize 2004  & \textcolor{red}{\scriptsize 2008} & \scriptsize 2012 & \scriptsize 2016 & \scriptsize 2005-2007 \\
		\cline{2-7}	
		&\scriptsize Predicted winning party & \scriptsize D & \textcolor{red}{\scriptsize D} & \scriptsize D & \scriptsize D & \scriptsize D \\
		\cdashline{2-7}[0.8pt/2pt]
		&\scriptsize Forecast accuracy in swing states & \scriptsize 84.6\% & \textcolor{red}{\scriptsize 100.0\%} & \scriptsize 100.0\% & \scriptsize 100.0\%  & \scriptsize 92.3\% \\
		\cdashline{1-7}[0.8pt/2pt]
		\scriptsize Number of swing states: 13 &\scriptsize Forecast accuracy in other states & \scriptsize 94.7\% & \textcolor{red}{\scriptsize 97.4\%} & \scriptsize 97.4\% & \scriptsize 97.4\% & \scriptsize 94.7\% \\
		\hline
		
		\multirow{3}*{\scriptsize The 2012 US Presidential Election} & \scriptsize Network & \scriptsize 2004  & \scriptsize 2008 & \textcolor{red}{\scriptsize 2012} & \scriptsize 2016 & \scriptsize 2009-2011 \\
		\cline{2-7}	
		&\scriptsize Predicted winning party & \scriptsize D & \scriptsize R & \textcolor{red}{\scriptsize D} & \scriptsize D & \scriptsize D \\
		\cdashline{2-7}[0.8pt/2pt]
		&\scriptsize Forecast accuracy in swing states &  \scriptsize 55.6\% & \scriptsize - & \textcolor{red}{\scriptsize 66.7\%} & \scriptsize 55.6\% & \scriptsize 55.6\% \\
		\cdashline{1-7}[0.8pt/2pt]
		\scriptsize Number of swing states: 9 &\scriptsize Forecast accuracy in other states & \scriptsize 92.9\% & \scriptsize - & \textcolor{red}{\scriptsize 100.0\%} & \scriptsize 97.0\% & \scriptsize 95.2\% \\
		\hline
		
		\multirow{3}*{\scriptsize The 2016 US Presidential Election} & \scriptsize Network & \scriptsize 2004  & \scriptsize 2008 & \scriptsize 2012 & \textcolor{red}{\scriptsize 2016} & \scriptsize 2013-2015 \\
		\cline{2-7}	
		&\scriptsize Predicted winning party & \scriptsize R & \scriptsize R & \scriptsize R & \textcolor{red}{\scriptsize R} & \scriptsize R \\
		\cdashline{2-7}[0.8pt/2pt]
		&\scriptsize Forecast accuracy in swing states &  \scriptsize 81.8\% & \scriptsize 72.7\% & \scriptsize 72.7\% & \textcolor{red}{\scriptsize 81.8\%} & \scriptsize 81.8\% \\
		\cdashline{1-7}[0.8pt/2pt]
		\scriptsize Number of swing states: 11 &\scriptsize Forecast accuracy in other states & \scriptsize 92.5\% & \scriptsize 100.0\% & \scriptsize 100.0\% & \textcolor{red}{\scriptsize 100.0\%} & \scriptsize 100.0\% \\
		
		\hline
		\hline 
	\end{tabular}
	
	\textit{Note: in the part of ‘Predicted winning party’, ‘R’ represents the Republican Party and ‘D’ represents the Democratic Party. If the predicted winning party does not match the fact, the forecast accuracy in swing states and that in other states will no longer be calculated. The cases producing the best prediction result are marked in red.}
\end{table}

Economic voting theory assumes that on an individual level voters react to economic indicators to hold incumbents responsible for the performance of the economy \citep{bengtsson2004economic}. The GDP data of the election year and its vicinity directly reflect the economic development at that time, which not only affects the political attitude of voters, but also reflects the influence of political policy on the economic field. This is consistent with the conclusion in paper \citep{dassonneville2017economic}; There is an association between economic indicators and levels of volatility since voters have to switch parties if they want to punish or reward political actors. As a result, based on the close relationship between politics and economy, the forecast results obtained by those data can better reflect reality and thus obtain higher accuracy.

\section{Case Analysis}

\subsection{Network Structure}

\quad\  Signed networks can be regarded as a special type of two-layer networks where the same set of nodes are connected by two logically contradictory types of links, so called positive and negative edges \citep{wang2019self}. Therefore, we split each network, which we generate to describe the relationships between the industrial structure of American states, into two weighted unsigned subnetworks, marked as the positive subnetwork and the negative subnetwork, and study the properties of each subnetwork.

\begin{table}[h]  
	\caption{ Network features of the signed economic networks.}  
	\label{tab4}
	\begin{tabular}{p{1cm}p{1.5cm}p{3cm}p{1.5cm}p{2.5cm}p{1.5cm}p{2cm}p{2.5cm}}
		\hline
		\hline
		\scriptsize \centering \textbf{Year} &\scriptsize \centering \textbf{Number of edges}& \scriptsize \centering \textbf{Proportion of Negative edges}& \scriptsize \centering \textbf{Mean of $w$ }& \scriptsize \centering \textbf{Standard deviations of $w$}& \scriptsize \centering \textbf{Subnetwork}& \scriptsize \centering \textbf{Average degree}& \scriptsize \textbf{\centering Average clustering coefficient}  \\
		\hline
	
		\multirow{2}*{\scriptsize \textbf{2008}} &\multirow{2}*{\scriptsize \centering 1214}& \multirow{2}*{\scriptsize \centering 0.3583}& \multirow{2}*{\scriptsize \centering 0.72} & \multirow{2}*{\scriptsize 0.28}& \scriptsize \textbf{Positive}& \scriptsize 31.16 & \scriptsize {\centering 0.866}  \\
		\cdashline{6-8}[0.8pt/2pt]
		& & & & &\scriptsize \textbf{Negative}& \scriptsize 17.06 & \scriptsize {\centering 0.620}  \\
		\cdashline{1-8}[0.8pt/2pt]
		
		\multirow{2}*{\scriptsize \textbf{2012}} &\multirow{2}*{\scriptsize \centering 1226}& \multirow{2}*{\scriptsize \centering 0.3564}& \multirow{2}*{\scriptsize \centering 0.74} & \multirow{2}*{\scriptsize 0.26}& \scriptsize \textbf{Positive}& \scriptsize 31.56 & \scriptsize {\centering 0.864}  \\
		\cdashline{6-8}[0.8pt/2pt]
		& & & & &\scriptsize \textbf{Negative}& \scriptsize 17.06 & \scriptsize {\centering 0.620}  \\
		\cdashline{1-8}[0.8pt/2pt]
		
		\multirow{2}*{\scriptsize \textbf{2016}} &\multirow{2}*{\scriptsize \centering 1225}& \multirow{2}*{\scriptsize \centering 0.3583}& \multirow{2}*{\scriptsize \centering 0.80} & \multirow{2}*{\scriptsize 0.19}& \scriptsize \textbf{Positive}& \scriptsize 32.00 & \scriptsize {\centering 0.846}  \\
		\cdashline{6-8}[0.8pt/2pt]
		& & & & &\scriptsize \textbf{Negative}& \scriptsize 16.67 & \scriptsize {\centering 0.627}  \\
		
		\hline
		\hline 
	\end{tabular}
	
	\textit{Note: '$w$' represents the modified cosine similarities between nodes.}
\end{table}

\begin{figure}[h]
	\centerline{\includegraphics[scale=0.22]{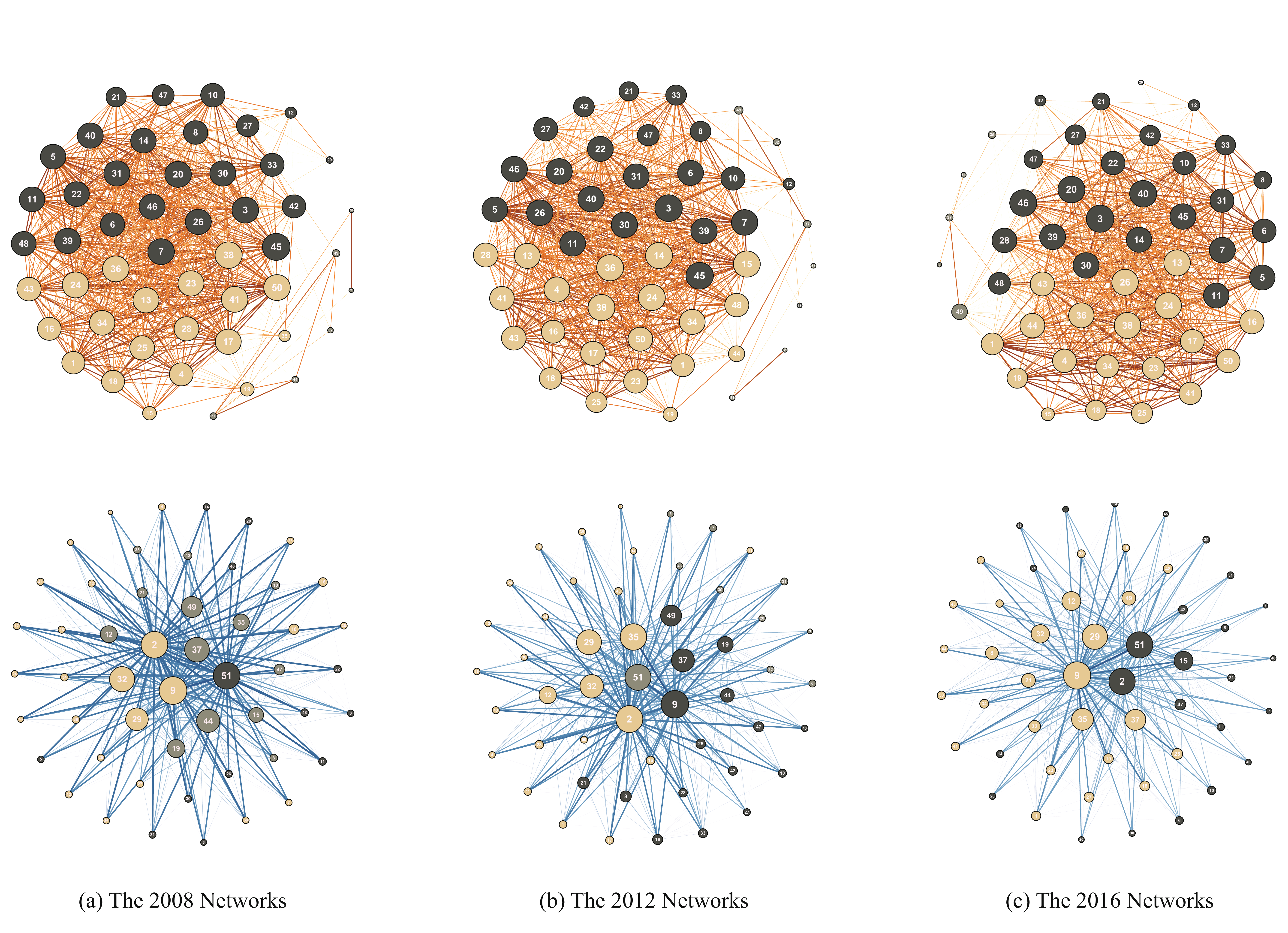}}
	\caption{Diagram of splitting each network of election years into a positive subnetwork and a negative subnetwork. The edges in positive subnetworks are drawn in orange, while those in negative networks are blue. The color of nodes reveals the community partition; nodes of the same color belong to the same community.} 
	\label{fig5}
\end{figure}

According to the principle of constructing our networks, the positive subnetworks record the similarity relationship higher than the mean, and the higher the edge weight is, the higher the similarity between the pair of nodes is. In contrast, the negative subnetworks record the lower part, and the higher the edge weight is, the lower the similarity is. Nevertheless, both in positive and negative subnetworks, when the edge weight is close to zero, it indicates that the similarity between the two nodes is at a medium level. The Fig. \ref{fig5} demonstrates the positive and the negative subnetworks of election years as an example. We can see that the negative network shows a star network with a small number of core nodes connected tightly, while the positive network shows that most of the nodes have dense connections, but a few nodes are connected sparsely at the periphery.

~\\
\hspace*{1em} 1) Macro-level Analysis

\begin{figure}
	\centerline{\includegraphics[scale=0.3]{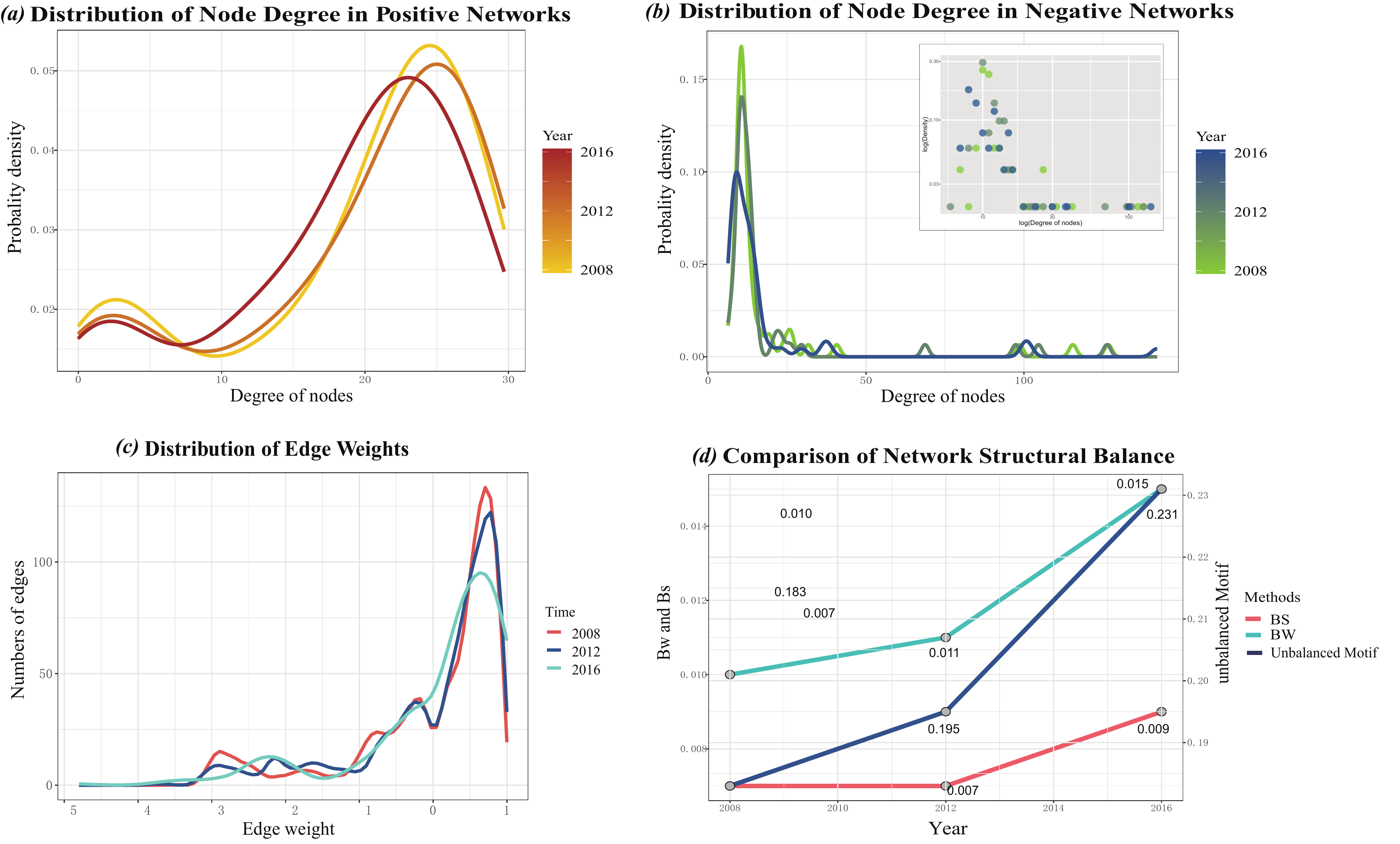}}
	\caption{The distribution of node degree in positive and negative subnetworks, the distribution of edge weights and the network structural balance of networks for election years between 2008 and 2016, shown from (a) to (d) respectively.} 
	\label{fig6}
\end{figure}

Taking into account the degree distribution, we find that the degree distribution of positive subnetworks presents a bimodal state, where the peak at high degree is much higher than that at low degree, but that of negative subnetworks is similar to the power law, shown in Fig. \ref{fig6}(a) and Fig. \ref{fig6}(b). Some nodes only appear in positive or negative subnetworks, for example, node No.9 only appears in the negative network. This phenomenon reveals that there is a high similarity between the industrial structures of most states, but there are also some states with special industrial structure, which are quite different from other states, such as Alaska and Wyoming. 

We also observe the distribution of edge weights. The range of node degree in the signed network widens from 2008 to 2016, but the overall proportion of negative edges decreases, shown in fig. \ref{fig6}(c) and Table \ref{tab4}, indicating that the development of individual states becomes increasingly idiosyncratic, such as Alaska.

In addition, the community partition of these networks based on the classical community discovery algorithm from Louvain \citep{blondel2008fast}, shown in Fig. \ref{fig5}, is highly similar to the real political camps, which once again proves the relationship between economy and politics. To observe the change of network structural balance, we used three algorithms to calculate: strong balance detection index $B_{Z}$ and weak balance detection index $B_{W}$ mentioned in paper \citep{kirkley2019balance}, and the method to calculate the proportion of unbalance motifs \citep{lewis2014us}. The results shows that the network structural balance grew up during the last three US presidential elections, shown in Fig. \ref{fig6}(d).

~\\
\hspace*{1em} 2) Micro-level Analysis

In order to study the role of each node in the network in detail, we choose the closeness centrality as the main evaluating indicator, which can measure the average distance between the target node and others, so that we can find the nodes that steers the direction of the dynamic evolution.

In the negative subnetworks, the closeness centralities of some swing states that changed from supporting the Democratic Party to the Republican Party in 2008-2016 decrease, while those of some swing states insisting on supporting the Democratic Party increase. The phenomenon is opposite in the positive subnetworks.

If we take into account the states with the lowest closeness centrality, it will be found that the dark-blue state District of Columbia and the dark-red state Wyoming have been at the core of the negative subnetworks and the positive subnetworks respectively for a long time, and the dark-red state Alabama is consistently at the core of both the positive and the negative subnetworks. Besides that, comparing the political tendencies of the top-ten states with lowest closeness centrality, the number of red states win slightly in both positive and negative subnetworks.

The above research on node centrality analyses the characteristics of some typical nodes in the network from a micro perspective, which helps us to observe the function of these states during the election.

\subsection{Economic Characteristics}

\begin{figure}
	\centerline{\includegraphics[scale=0.3]{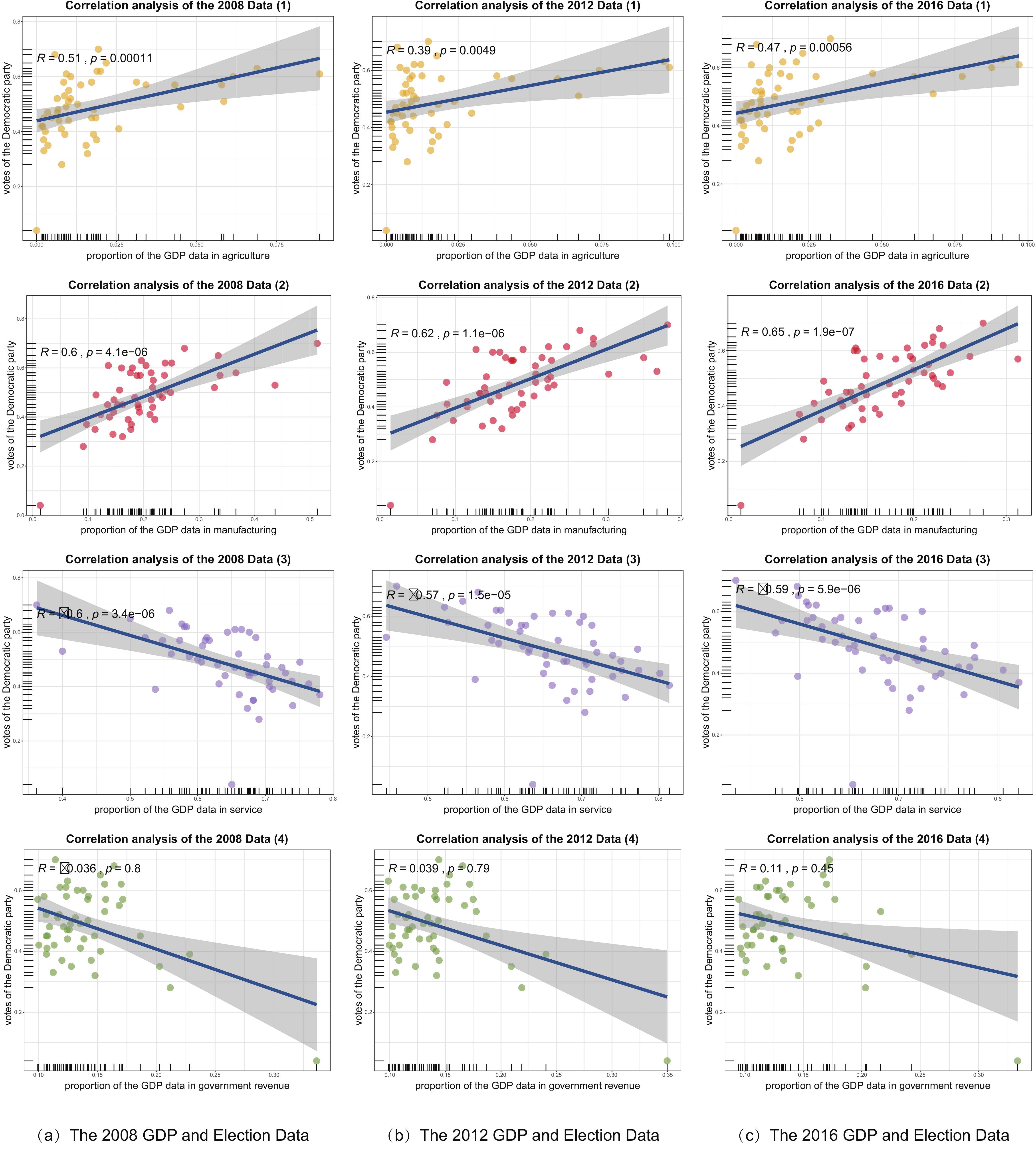}}
	\caption{The Spearman Correlation Analysis between the vote of the Democratic party and the market share of each industry.} 
	\label{fig7}
\end{figure}

\quad\  Apart from the network structure, we also analyzed the changes in economic characteristics. We first use the Spearman Correlation Analysis to study the relationship between the vote of each political party and the market share of each industry. The result shows that there is a positive correlation between the vote of the Democratic Party and the proportion of the third industry, shown in Fig. \ref{fig7}, while the vote of the Republican Party is positively related to the proportion of the second industry. This demonstrates that considering different political parties have their own economic strategies, regions with different industrial structure characteristics have unique biases towards political parties.

The states who changed the political parties that they supported during 2000-2016 are also collected, and the Mann-Whitney tests on them show that there is a significant difference in the proportion of varies industries between states turning blue and states turning red, especially in government revenue. According to \citet{acemoglu2005politics}, government revenues as a fraction of GDP appear to be higher in richer countries and in societies that are generally considered to have more “constrained” governments, therefore our results confirms that the two political parties have different governing ideas. In addition, the mean value of cosine similarity rises and its standard deviation shows a downward trend during the last three elections, shown in Table \ref{tab4}. This reveals an increase in the similarity of industrial structure among states.

\section{Conclusion}

\quad\  It is a valuable future research direction to utilize the relationships between systems to help researchers get better understanding of individual system. In this work, expanding beyond the previous narrow view, we study the political system and economic system from a systematic perspective, and treat them as a complex whole to explore the coupling interaction between these two systems. 

We build association networks of industrial structure between different regions based on GDP data, and study the diffusion and evolution of opinions during the US presidential election through a kinetic model in statistical physics called the Potts Model. In this way, our work proposes a simple and efficient prediction model for the US presidential election, and demonstrates the close relationship between economic structure and political attitude in the United States with the help of networks extracted from economic data. The analysis of the network structure and economic characteristic supplements the specific features of economic development during this period and the interaction between political tendency and industrial structure.

In addition to the election forecasting, considering that the model setup can indicate the political impact of industrial development patterns of political individuals, we think it may be useful to model the economic structure, and may help to evaluate the economic policies and adjust the development direction. This requires further exploration in the future.

%\quad 1)	$p\_p=0.1$ and $p\_n=0.1$, that is, positive and negative edges exist within and between communities. (To illustrate this network structure, we used 16 nodes as an example to make schematic diagram, shown in Fig. 2(b) BMs.)

%1) \textit{(Critical Ratio)} There is a critical negative edge ratio, $\sigma$. \textcolor{red}{The critical ratio stands for a turning point, which means, when the proportion of negative edges in the network exceeds the critical ratio, there appear to be two distinct evolutionary results. The detection of the critical ratio depends on a change-points detection method, pettitt, which is a nonparametric test method [21].}  

%\textcolor{red}{What’s more, we also observe that even if the evolution results of proportion look the seem, the distributions of nodes states are quite different in signed networks with different structures. The stricter the community structure of}

%
\subsubsection{Acknowledgements}
\quad\  This work is supported by the National Natural Science Foundation of China (Grant Nos. 61573065 and 71731002) and BNU Interdisciplinary Research Foundation for the First-Year Doctoral Candidates (Grant No. BNUXKJC1921).

\renewcommand\bibname{References}
\bibliographystyle{elsarticle-harv}  
\bibliography{read-reference}                  

\end{document}